Weak antilocalization induced by Se substitution in layered BiCh$_2$-based (Ch = S, Se) superconductors LaO$_{1-x}$F$_x$BiS$_{2-y}$Se$_y$

K. Hoshi[1], H. Arima[1], N. Kataoka[2], M. Ochi[3,4], A. Yamashita[1], A. de Visser[5], T. Yokoya[2,6], K. Kuroki[4] and Y. Mizuguchi[1]*

[1]Department of Physics, Tokyo Metropolitan University, Hachioji 192-0397, Japan
[2]Graduate School of Natural Science and Technology, Okayama University, Okayama 700-8530, Japan
[3]Forefront Research Center, Osaka University, Toyonaka, Osaka 560-0043, Japan
[4]Department of Physics, Osaka University, Toyonaka, Osaka 560-0043, Japan
[5]Van der Waals–Zeeman Institute, University of Amsterdam, Science Park 904, 1098 XH Amsterdam, The Netherlands
[6]Research Institute for Interdisciplinary Science, Okayama University, Okayama 700-8530, Japan

**Abstract**

We report transport properties for layered BiCh$_2$-based (Ch = S, Se) superconductors LaO$_{1-x}$F$_x$BiS$_{2-y}$Se$_y$ ($x$ = 0.2, 0.5, $y$ = 0–1.05) and the observation of *weak antilocalization* (WAL). Electrical resistivity and Hall coefficients for the Se-poor samples increase with decreasing temperature. The increase becomes less pronounced with increasing Se concentration indicating a loss of insulating behavior. Interestingly, the moderately Se-substituted samples exhibit metallic behavior in the high-temperature region and a weak increase in the resistivity in the low-temperature regions, which indicates the existence of carrier localization. The heavily Se-substituted compounds show metallic behavior in the entire-temperature region. Sign changes of the Hall coefficients are observed for the $x$ = 0.2 samples, which possibly is related to a charge-density wave (CDW). Magnetoresistance measurements indicate that WAL is realized in the heavily Se-substituted systems. The WAL behavior is weakened by the changes in F and Se concentrations. A crossover state of the WAL and WL emerges around the moderately F-doped and Se-free LaO$_{0.8}$F$_{0.2}$BiS$_2$. The change of the resistivity behavior by the F and Se substitution clearly correlates to the difference of the magnetoconductance. Moreover, the localization regions of the WAL-WL crossover and *weak* WAL states are possibly associated with the CDW. We propose that the BiCh$_2$-based system is a good platform for studying relationship between WAL, superconductivity, and electronic ordering because those states are tunable by element substitutions with bulk single crystals.

**Significance**

Weak antilocalization (WAL) is an important quantum phenomenon, and strong spin-orbit coupling is essential for WAL states to emerge. Studies of WAL largely focus on two-dimensional materials and thus WAL research on bulk single crystals is limited. Here, we find that typical WAL signatures, such as sharp cusps in the low-field magnetoresistance, are observed for bulk single crystals of layered $BiCh_2$-based superconductors. The WAL states are gradually suppressed and a crossover of WAL and weak localization is induced by element substitution. Moreover, the change of the magnetoconductance clearly correlates with the modification of the resistivity and Hall coefficient by elemental substitution. Our findings suggest that $BiCh_2$-based superconductors are good platforms for investigating the relationship between localization and superconductivity.

**Introduction**

Weak antilocalization (WAL) has been extensively studied in topological-materials and spintronics [1–6]. Strong spin-orbit coupling (SOC) basically leads to WAL [7, 8]. A crossover from WAL to weak localization (WL) may be induced by changing parameters, such as impurity, elemental-substitution, and carrier concentration [1, 3, 5]. Two-dimensional (2D) materials have been the central target for the WAL and WL. Here, we propose that the layered $BiCh_2$-based (Ch = S, Se) superconductors are novel candidate system for studying the WAL and WL in spite of being bulk single crystals.

$BiCh_2$-based superconductors have a layered crystal structure composed of blocking layers and $BiCh_2$ conducting layers [9, 10]. The typical $BiCh_2$-based system is $REOBiCh_2$-type (RE = Rare earth) [Fig. 1(a)] and the parent compound is a band insulator [10–14]. F substitution for O sites leads to electron-carrier doping, and superconductivity emerges at low temperatures (the superconducting transition temperature $T_c$ ~ 2–10 K) [10–14], which is in a certain way similar to the situation in iron-based superconductors REOFeAs [15]. The Bi-$6p_x/p_y$ orbitals largely contribute to the electronic properties since the conduction bands are mainly composed of these orbitals [16–18]. For the $REO_{1-x}F_xBiCh_2$-type system, rectangular electron pockets are found at the $X$ points up to $x$ ~ 0.45 [19–21]. By increasing $x$, a Lifshitz transition takes place around $x$ ~ 0.45, and hole pockets emerge around the $\Gamma$ and $M$ points [20, 21].

As regards the superconducting state, theoretical calculations suggest that an unconventional Cooper pairing mechanism and anisotropic superconducting gap structures can be realized in the $BiCh_2$-based system [16, 18]. However, many experimental results, such as the magnetic penetration depth, specific heat, and thermal conductivity, provided evidence that superconductivity is of the conventional isotropic $s$-wave type [22–25]. The residual resistivity is in the order of m$\Omega$cm, even in single crystalline samples, which indicates that superconductivity occurs in the dirty limit. The robustness against impurities can support the conventional isotropic *s*-wave scenario. In contrast, in an angle-resolved photoemission spectroscopy (ARPES) study an anisotropic superconducting gap with nodelike minima was directly observed, which suggested that anisotropic *s*-wave superconductivity with accidental nodes can be realized in the $BiCh_2$-based system [26]. Moreover, the isotope effect exponent $\alpha$ in the $BiCh_2$-based systems strikingly deviates from the standard BCS value of $\alpha \sim 0.5$ which possibly indicates an unconventional Cooper pairing mechanism [27, 28]. A checkerboard stripe pattern observed by scanning tunneling microscopy/spectroscopy (STM/STS) may as well be related to an unconventional pairing mechanism [29, 30]. A relevant feature of the tetragonal crystal structure of the $BiCh_2$-based compounds (space group *P*4/*nmm*, No. 129) is that it has local inversion symmetry breaking in the $BiCh_2$ layer, which can cause a spin polarization on the Fermi surface (FS) due to the local Rashba-type SOC [31]. A spin- and angle-resolved photoemission spectroscopy (SARPES) study directly found a Rashba-like spin texture near the Fermi level [32]. Furthermore, the unique spin texture can lead to the protection of the Cooper pairs from depairing due to a magnetic field, and a suppression of the paramagnetic pair-breaking effect. More recently, extremely high upper critical fields were reported for $LaO_{0.5}F_{0.5}BiS_{2-y}Se_y$ ($y$ = 0.22 and 0.69), which indicates that the Rashba-type spin texture plays a significant role in superconductivity [33]. This provides strong evidence that an unconventional superconducting state can be realized in the $BiCh_2$-based system.

Although the superconducting properties have been intensively investigated in the $BiCh_2$-based compounds, there is an unresolved problem in the normal-state-resistivity behavior. In previous experimental results on $LaO_{0.5}F_{0.5}BiS_2$, a weak increase of the resistivity was observed, although superconductivity emerges at low temperature [11, 34]. The experimental results are not consistent with the theoretical calculations since the Fermi level crosses the conduction band for the doping level $x = 0.5$ and thus metallic behavior should emerge [18, 19]. Furthermore, the shielding volume fraction estimated from magnetization measurements is less than 10%, which indicates superconductivity in $LaO_{0.5}F_{0.5}BiS_2$ is filamentary in nature [10, 11, and *SI Appendix,* Fig. S3]. Se substitution leads to the gradual suppression of the upward behavior in the resistivity and heavily Se-doped $LaO_{0.5}F_{0.5}BiSSe$ shows almost metallic behavior and bulk superconductivity (the shielding fraction is roughly 100%) [10, 34, and *SI Appendix,* Fig. S3]. The Se-substitution effect is considered to act as chemical pressure, which leads to the depression of the in-plane disorder [10, 35]. These results imply that a localization phenomenon is in play that can be suppressed by Se substitution. Sakai *et al*. suggested that similar localized states, reported for $Sr_{1-x}La_xFBiS_2$, which is closely related to $LaO_{1-x}F_xBiS_2$, may be due to Anderson localization [36]. If the states are related to Anderson localization, a negative magnetoresistance (MR) is expected. However, MR data were not reported in the previous work. This justifies a systematic study to address the localization phenomena in $BiCh_2$-based systems by means of magnetotransport.

Here, we report the transport properties of single crystalline $LaO_{1-x}F_xBiS_{2-y}Se_y$ ($x$ = 0.2, 0.5, $y$ = 0–

1.05). An insulating behavior is observed for moderately F-doped and Se-free $LaO_{0.8}F_{0.2}BiS_2$. Heavily F-doped $LaO_{0.5}F_{0.5}BiS_2$ shows superconductivity at low temperature although a weak increase in the resistance is still observed. The increase is progressively depressed with increasing Se concentration for both $x = 0.2$ and 0.5. The moderately Se-substituted samples exhibit metallic behavior in the high-temperature region and superconductivity at low temperatures. The weak increase in the resistivity still exists at low-temperatures, which implies the existence of localized states. Metallic behavior in the entire temperature region emerges in the heavily Se-substituted compounds for both $x = 0.2$ and 0.5. The Hall coefficients in the low-temperature region are also depressed with increasing Se concentration, in accordance with the resistivity data. Moreover, sign changes of the Hall coefficients are observed for $x = 0.2$, which may relate to a charge-density wave (CDW). To further investigate the localization state, we performed MR measurements. Sharp cusps of the MR in the low-field regions are observed in the heavily Se-substituted compounds, which indicates the WAL state emerge. The WAL is weakened for the moderately Se-substituted compounds, as well as for heavily F-doped and Se-free $LaO_{0.5}F_{0.5}BiS_2$ (we call this *weak* WAL in the following). Furthermore, a WAL-WL crossover is found in moderately F-doped and Se-free $LaO_{0.8}F_{0.2}BiS_2$. The experimental results are almost consistent with a theoretical prediction which suggests that a crossover from WAL to WL occurs when the Fermi energy or band gap changes [8]. The electronic band structures from the theoretical calculations and ARPES spectra clearly show that Se substitution leads to decreasing the band gap. The change from the WAL to *weak* WAL and the WAL-WL crossover by the F and/or Se substitution obviously correlate with the results of the resistivity and Hall coefficient. Therefore, the elemental substitution of F and Se in the $BiCh_2$-based system leads to changing the localization states with controlling the Fermi energy and band gap. Moreover, it may be significant to suppress the local crystal disorder to cause the WAL states, given the previous studies. The localization states of the *weak* WAL and the WAL-WL crossover may be associated with the CDW.

**Results and discussion**

Figure 1(a) shows the crystal structure of $LaO_{1-x}F_xBiS_{2-y}Se_y$ which is a layered material composed of La(O/F) blocking layers and $BiCh_2$ conducting layers. The Se in this system is selectively substituted at the Ch1 sites [37]. Figures 1(b) and (c) present the temperature dependence of the electrical resistivity in the *ab*-plane $\rho_{ab}$ for $x = 0.2$ and $x = 0.5$, respectively. For $x = 0.2$ [Fig. 1(b)], Se-free $LaO_{0.8}F_{0.2}BiS_2$ ($x = 0.2$, $y = 0$) exhibits insulating behavior in the entire temperature range. The insulating behavior is gradually suppressed with increasing Se concentration, and superconductivity is observed at low temperatures for $y = 0.46$, 0.72, and 0.94. However, a weak increase of $\rho_{ab}$ in the low-temperature region is still seen for $y = 0.46$ and 0.72, while in the high-temperature region metallic behavior is observed. When the Se-substitution level reaches $y = 0.94$, metallic behavior is found in the entire temperature range. The $x = 0.5$ samples [Fig. 1 (c)] show a trend similar to the $x = 0.2$ samples. However, there is a clear difference between the Se-free samples: $LaO_{0.5}F_{0.5}BiS_2$ ($x = 0.5$, $y = 0$) shows a superconducting transition at $T \sim 2$ K, while $LaO_{0.8}F_{0.2}BiS_2$ ($x = 0.2$, $y = 0$) is insulating behavior in the entire temperature range down to 2 K. In addition, the absolute value of $\rho_{ab}$ at low temperatures for $LaO_{0.5}F_{0.5}BiS_2$ is much smaller than that of $LaO_{0.8}F_{0.2}BiS_2$, despite the weakly insulating behavior for $LaO_{0.5}F_{0.5}BiS_2$. These results indicate that electron-carrier doping (F substitution) suppresses the insulating characteristics, but a F content $x = 0.2$ is not sufficient to suppress the insulating state completely. However, heavily Se-substituted $LaO_{0.8}F_{0.2}BiS_{1.06}Se_{0.94}$ ($x = 0.2$, $y = 0.94$) with the same nominal $x = 0.2$ shows metallic behavior and superconductivity. Moreover, theoretical calculations suggest that a doping level of $x = 0.2$ should be sufficient to make the BiCh$_2$-based system metallic [18, 19]. The moderately Se-substituted samples ($y = 0.46$ and $0.72$ with $x = 0.2$ and $y = 0.39$ and $0.65$ with $x = 0.5$) exhibit metallic behavior in the high-temperature region, but a weak increase in $\rho_{ab}$ at low-temperatures. Hence, we can deduce that carrier localization occurs in these compositions. Thus, the $LaO_{0.8}F_{0.2}BiS_2$ phase is not a conventional band insulator but an insulator caused by carrier localization. A similar localization state was observed for polycrystalline samples of $Sr_{1-x}La_xFBiS_2$, which compares well to our target material $LaO_{1-x}F_xBiS_{2-y}Se_y$ [36]. The authors of Ref. 36 suggested that the weak increase in $\rho_{ab}$ is due to Anderson localization. We will discuss this possibility in the section with the MR results.

To further investigate the physical properties in the normal states, we performed Hall resistivity measurements. Figures 1(d) and (e) show the temperature dependence of the Hall coefficient $R_H$ for $x = 0.2$ and $x = 0.5$, respectively. As shown in the inset of Fig. 1(e), our experimental results show single-band behavior, i.e. a linear function $\rho_H(B)$, although multiband-like behavior [nonlinear $\rho_H(B)$] was reported for BiCh$_2$-based compounds in previous studies [38, 39]. Note that in the present work we use single crystals, while in the previous experiments [38, 39] the data were obtained on polycrystalline samples, which might cause some differences. We determined the Hall coefficient $R_H$ from the linear slope of $\rho_H (B)$. $R_H$ increases with decreasing temperature for $x = 0.2$ and $0.5$. The increasing behavior, is systematically depressed by increasing Se concentration, analogous to the $\rho_{ab}$ ($T$) behavior in Figs. 1(b) and (c). For the $x = 0.2$ samples [Fig. 1(d)], sign changes of $R_H$ from negative to positive are observed for $y = 0$, $0.24$, $0.47$, and $0.72$. In contrast, a negative $R_H$ is kept until 5 K for heavily Se-substituted $LaO_{0.8}F_{0.2}BiS_{0.95}Se_{1.05}$ ($x = 0.2$, $y = 1.05$), which indicates that electron-type carriers are dominant in the entire temperature region for this compound. For the $x = 0.5$ samples [Fig. 1(e)], the $R_H$ ($T$) curves are similar to those for $x = 0.2$. However, heavily Se-substituted $LaO_{0.5}F_{0.5}BiS_{1.09}Se_{0.91}$ ($x = 0.5$, $y = 0.91$) shows a positive sign of $R_H$ throughout the entire temperature region, which implies that hole-type carriers are predominant for that compound. According to the theoretical calculations for the FS topology of the BiCh$_2$-based system, the electron pockets around the $X$ points exist up to $x \sim 0.45$, and hole pockets around the $\Gamma$ and $M$ points emerge above $x > 0.45$ [20, 21]. Therefore, the negative $R_H$ (electron-type carriers) for $LaO_{0.8}F_{0.2}BiS_{0.95}Se_{1.05}$ and the positive $R_H$ (hole-type carriers) for $x = 0.5$ in the low-temperature region are in line with the theoretical work. However, the sign changes and the hole-type carriers at low temperatures observed for $y = 0$–0.72 in $x = 0.2$ are not consistent with the theoretical expectations. A similar sign-change behavior of $R_H$ was observed in polycrystalline $EuFBiS_2$, which was attributed to a CDW [40]. In our experiments, the sign changes in $R_H$ are observed concomitant with the upward behavior of $\rho_{ab}$ ($T$), which is incompatible with a conventional CDW scenario by FS nesting. However, the nonmetallic behavior probably does not originate from a conventional band insulator, since superconductivity is observed even when $\rho_{ab}$ ($T$) increases at low temperatures. In addition, a recent synchrotron radiation X-ray diffraction study suggested the formation of a CDW below $T \sim 260$ K in single crystals of $LaO_{0.5}F_{0.5}BiS_2$ [41]. Besides there are several theoretical papers in which a CDW for the $BiCh_2$-based compounds is suggested [42, 43]. Thus, the sign changes in $R_H$ for our crystals may relate to the CDW formation. Moreover, the sign-change temperature progressively decrease, and the values of the positive $R_H$ in the low-temperature regions are also systematically depressed with increasing Se concentration. Therefore, together with the results of the $\rho_{ab}$ ($T$) measurements, we speculate that the $R_H$ data point to carrier localization in the Se-free and moderately Se-substituted compounds and that this may be linked to the CDW. The indication of the hole-type carriers in the low-temperature regions for the $y = 0$–0.72 in the $x = 0.2$ results from the localization where a negative $R_H$ should result from the carrier doping level. For $x = 0.5$, the sign-change behavior is observed only for $y = 0.19$. It may be difficult to observe the sign change in the $x = 0.5$ case, since basically, the hole-type carriers are predominant in the carrier-doping level. Alternatively, the temperature dependence of the Hall coefficients for $x = 0.5$ has almost the same tendency with increasing Se concentration as $x = 0.2$. Therefore, we expect that the CDW might also be present in the $x = 0.5$ samples. The carrier density $n$ at low temperatures, estimated from $n = 1/|R_H e|$, changes from the order of $10^{19}$ cm$^{-3}$ to $10^{21}$ cm$^{-3}$ by Se substitution in $x = 0.2$ and from the order of $10^{19}$ cm$^{-3}$ to $10^{20}$ cm$^{-3}$ in $x = 0.5$.

To further investigate the localization behavior deduced from $\rho_{ab}$ ($T$) and $R_H$ ($T$), we performed MR measurements. If Anderson localization is realized in the $BiCh_2$-based systems as suggested in a previous study [36], a negative MR should be observed in the localization regime [44, 45]. Figures 2 (a)–(d) show the MR defined as MR = [$\rho_{ab}$ ($B$) − $\rho_{ab}$ (0)]/$\rho_{ab}$ (0) × 100% as a function of magnetic field for $y = 0$ and 1.05 with $x = 0.2$, and $y = 0$ and 0.91 with $x = 0.5$, at $T = 5$ K and 7.5 K, respectively. The observed MR is clearly different from the negative magnetoresistance expected from the Anderson localization scenario. The MR curves for the heavily Se-substituted compounds in $x = 0.2$ and 0.5 show sharp cusps in low magnetic field at $T = 5$ K. The sharp dip of the MR in low field is a characteristic feature of WAL which has been observed in various materials with strong SOC [1–4]. This is consistent with the previous ARPES observation that SOC gives a pronounced effect on the band structure of $La(O,F)BiS_2$ [46]. The low-field MR curves for heavily F-doped and Se-free $LaO_{0.5}F_{0.5}BiS_2$ become broad and close to classical quadratic in field behavior. Moderately F-doped and Se-free $LaO_{0.8}F_{0.2}BiS_2$ exhibit an almost flat MR in low fields. By increasing the applied field, a maximum around $B = 3$ T is observed, resulting in a negative MR at even higher fields. This indicates that the WL state emerges in the high-field region. At $T = 7.5$ K [Figs. 2(c) and (d)], the WAL for the heavily Se-substituted samples with $x = 0.2$ and 0.5 is weakened, and the WL behavior in the high fields becomes clear for moderately F-doped and Se-free $LaO_{0.8}F_{0.2}BiS_2$. To clearly exhibit the WAL, we correct the MR to the magnetoconductance $\Delta G$ (MC) defined as the $\Delta G(B) = G(B) - G(0)$. Figures 2(e)–(h) show the low-field MC as a function of magnetic field at $T = 5$ K [Figs. 2(e) and (f)] and 7.5 K [Figs. 2(g) and (h)]. We just show positive-field regions to analyze the MC data. The MR curves for moderately Se-substituted $LaO_{0.5}F_{0.5}BiS_{1.35}Se_{0.65}$ ($x = 0.5$, $y = 0.65$) and $LaO_{0.8}F_{0.2}BiS_{1.28}Se_{0.72}$ ($x = 0.2$, $y = 0.72$) are displayed in *SI Appendix*, Fig. S2. At $T = 5$ K, the MC curves for the heavily Se-substituted $LaO_{0.5}F_{0.5}BiS_{1.09}Se_{0.91}$ and $LaO_{0.8}F_{0.2}BiS_{0.95}Se_{1.05}$ steeply decrease with increasing magnetic fields. The negative MC is still observed for moderately Se-substituted $LaO_{0.8}F_{0.2}BiS_{1.28}Se_{0.72}$, $LaO_{0.5}F_{0.5}BiS_{1.35}Se_{0.65}$, and also heavily F-doped and Se-free $LaO_{0.5}F_{0.5}BiS_2$, but the decline of the MC is gradually suppressed with decreasing Se concentration. The MC for moderately F-doped and Se-free $LaO_{0.8}F_{0.2}BiS_2$ is almost independent of the magnetic field. The MC curves at $T = 7.5$ K have a similar trend as those at $T = 5$ K. The experimental results indicate that the heavily Se-substituted samples exhibit the WAL state, which is weakened in the moderately Se-substituted compounds and heavily F-doped and Se-free $La_{0.5}F_{0.5}BiS_2$. Furthermore, the almost flat MC behavior for moderately F-doped and Se-free $LaO_{0.8}F_{0.2}BiS_2$ implies a WAL-WL crossover takes place. Basically, the 2D MC of the WAL contribution is described by the Hikami-Larkin-Nagaoka (HLN) formula [7]. To quantitatively analyze the MC curves and compare them with previous studies of WAL, we fit the data with the 2D HLN formula written as

$$\Delta G = \frac{\alpha e^2}{2\pi^2 \hbar} \left[ \psi \left( \frac{1}{2} + \frac{\hbar}{4el_\phi^2 B} \right) - \ln \left( \frac{\hbar}{4el_\phi^2 B} \right) \right], \quad (1)$$

where $\alpha$ is prefactor, $l_\phi$ is the phase coherence length, and $\psi$ is the digamma function. The solid lines in Figs. 2(e)–(h) represent the HLN fitting. The MC data can be well fitted by the HLN model except for $LaO_{0.8}F_{0.2}BiS_2$ at $T = 7.5$ K [red solid line in Fig. 2(g)]. The coefficient of determination $R^2$ for this compound is less than 80% in spite of over 95% for the others [*SI Appendix*, Table S2]. This is probably due to the WAL-WL crossover state. The obtained values of $\alpha$ and $l_\phi$ are displayed in Figs. 4 (c) and (d). The magnitudes of $\alpha$ are in the order of $10^4$, except for $LaO_{0.8}F_{0.2}BiS_2$, which is much larger than reported for 2D compounds [1–4]. Such huge values of $\alpha$ have been observed in three-dimensional (3D) bulk systems and are considered to derive from the 3D bulk contribution [47–50]. Thus, we speculate that the large values of $\alpha$ in our systems are due to the 3D bulk nature of the single crystals. On the other hand, the obtained $l_\phi$ is comparable to the values found in previous studies for both 2D and 3D bulk compounds. The absolute values of $\alpha$ and $l_\phi$ systematically increase with increasing Se concentration [see Figs. 4(c) and (d)], which indicates that the WAL behavior emerges by Se substitution. Furthermore, the order of $\alpha$ for LaO$_{0.8}$F$_{0.2}$BiS$_2$ is $10^4$ times smaller than in the other samples, which implies that this compound is in the WAL-WL crossover states. A similar evolution of $\alpha$ by doping was observed for Bi$_{2-x}$Cr$_x$Se$_3$ and the sign-change behavior of $\alpha$ from negative to positive was regarded as the crossover from WAL to WL [3]. We analyzed the data with the HLN model so far, but a recent theoretical calculation about the WAL may be useful for our experimental results. The theoretical work which considered the strong SOC in the atoms that constitute the lattice (denoted as SOC lattice in Ref. 8) suggested that the WAL-WL crossover is realized by adjusting the ratio of the Fermi energy $E_F$ to the band gap $E_g$ [8]. In the BiCh$_2$-based system, several electronic-band calculations suggest that $E_g$ decreases by Se substitution [51, 52]. Since the F substitution causes electron-carrier doping, the elemental-substitution effect of Se and F leads to controlling $E_F$ and $E_g$ in the BiCh$_2$-based system, which contributes to the change of the MC. Moreover, the predicted $\Delta G$ ($B$) for the Bi$_{1-x}$Sb$_x$ alloy [8], in which $E_F$ of semi metallic Bi can be changed by Sb substitution, is alike the observed MC [Figs. 4(e)–(h)]. Thus, we expect that the change from the WAL to *weak* WAL and the WAL-WL crossover can be controlled by the elemental substitution of F and Se in the BiCh$_2$-based system and that the localization states observed in $\rho_{ab}$ ($T$) and $R_H$ ($T$) [Figs. 2 and 3] can be associated with the MC behavior.

To confirm the scenario on the changes of the MC behavior by controlling $E_F$ and $E_g$ by Se substitution, we consider the electronic band structures. Figures 3(a) and (b) exhibit the calculated electronic band structure for Se-free LaO$_{0.5}$F$_{0.5}$BiS$_2$ and heavily Se-substituted LaO$_{0.5}$F$_{0.5}$BiSSe. The band gap at the *X* point clearly decreases with Se substitution. Moreover, Figs. 3(c) and (d) show the ARPES spectra for our single crystals of Se-free LaO$_{0.5}$F$_{0.5}$BiS$_2$ and heavily Se-substituted LaO$_{0.5}$F$_{0.5}$BiSSe. The observed band gap values at the *X* point were estimated to be 0.875 eV for Se-free LaO$_{0.5}$F$_{0.5}$BiS$_2$ and 0.715 eV for heavily Se-substituted LaO$_{0.5}$F$_{0.5}$BiSSe. The tendency of the change of the band gap observed in ARPES is consistent with the theoretical calculations. In addition, the spectral intensity at $E_F$ of Se-free LaO$_{0.5}$F$_{0.5}$BiS$_2$ is strongly reduced compared to that of heavily Se-substituted LaO$_{0.5}$F$_{0.5}$BiSSe [see Fig. 3(e)], which is consistent with the localized and non-localized nature of the samples. Thus, the electronic band structure from the theoretical calculations and ARPES supports the results of the transport properties.

Next, we discuss a clear maximum of the MR for moderately F-doped and Se-free LaO$_{0.8}$F$_{0.2}$BiS$_2$. The theoretical calculation suggested that there is no minimum of $\Delta G$ ($B$) (no maximum of the MR) as a function of the field in the SOC lattice, while a clear minimum of the $\Delta G$ ($B$) (maximum of the MR) is observed in conventional HLN theory, which considers a 2D metal with strong spin-orbit coupled impurities [8]. We have not found any specific impurities in our single crystal samples (we confirmed that the actual atomic ratios are almost consistent with the nominal values although we cannot wholly determine that there are no impurities), but moderately F-doped and Se-free LaO$_{0.8}$F$_{0.2}$BiS$_2$ probably have disorder originating from local crystal distortions caused by structural instability due to the presence of Bi lone pairs [35]. The crystal structure of the parent compound has a monoclinic symmetry (space group $P2_1/m$) [53], and the tetragonal structure is gradually stabilized by the F substitution (electron-carrier doping) [54, 55]. Moreover, the Se-substitution effect causes the depression of the local disorder in the BiCh1 plane [35]. Thus, moderately F-doped and Se-free LaO$_{0.8}$F$_{0.2}$BiS$_2$ still have local distortions, which may lead to the clear maximum of the MR. On the basis of the discussion here, we propose that the BiCh$_2$-based LaO$_{1-x}$F$_x$BiS$_{2-y}$Se$_y$ compounds provide a new platform where the WAL states are realized when the structural disorder is removed by Se substitution.

In Figs. 4(a) and (b), we summarize the electronic phases of LaO$_{0.8}$F$_{0.2}$BiS$_{2-y}$Se$_y$ and LaO$_{0.5}$F$_{0.5}$BiS$_{2-y}$Se$_y$ as functions of $y$ and temperature. For $x = 0.2$ [Fig. 4(a)], the $\rho_{ab}$ ($T$) behavior for $y = 0$ demonstrates an insulating behavior in the entire temperature region [blue region in Fig. 4(a)]. The insulating behavior is gradually suppressed with increasing Se concentration, and superconductivity emerges for $y = 0.46$, 0.72, and 0.94, although the weak increase in $\rho_{ab}$ ($T$) is still observed for $y = 0.46$ and 0.72 (denoted as Weakly insulating in Fig. 4). There is no strong correlation between $y$ and $T_c$ in $x = 0.2$. When the Se-substitution level reaches $y = 0.94$, the $\rho_{ab}$ ($T$) curve exhibits the metallic behavior in the whole-temperature region (yellow regions in Fig. 4). The $T_{min}$ and $T^*$ [the definition is described in the caption of Fig. (4)] progressively decline with increasing Se concentration, and the sign change disappears for $y = 1.05$, which indicates that the localization state is suppressed by the Se substitution. The normalized resistivity $\rho_{ab}$ (5 K)/$\rho_{ab}$ (295 K) and the Hall coefficients at $T = 5$ K and 100 K systematically decrease with increasing Se concentration as shown in Figs. 4(c) and (d). The magnitude of $\alpha$ and $l_\phi$ in the HLN model increases with increasing Se concentration, which implies that the WAL state emerges for $y = 1.05$ and it is weakened for $y = 0.72$ [denoted as *Weak* WAL in Fig. 4]. Furthermore, the absolute value of $\alpha$ for Se-free LaO$_{0.8}$F$_{0.2}$BiS$_2$ is much smaller than the others, which indicates that the WAL-WL crossover state is realized. The electronic phase diagram for $x = 0.5$ [Fig. 4(b)] is similar to the one for $x = 0.2$; however, $y = 0$ in the $x = 0.5$ case presents superconductivity at $T \sim 2$ K, although $\rho_{ab}(T)$ exhibits a weakly insulating behavior. The largest absolute values of $\alpha$ and $l_\phi$ for $y = 0.91$ in the $x = 0.5$ samples imply the WAL characteristics [see Fig. 4(d)]. The WAL is diminished for LaO$_{0.5}$F$_{0.5}$BiS$_{1.35}$Se$_{0.65}$ and LaO$_{0.5}$F$_{0.5}$BiS$_2$. Therefore, the change of the MC behavior by the F and Se substitution is associated with the $\rho_{ab}$ ($T$) behavior for both $x = 0.2$ and 0.5. The $T_c$ for $x = 0.5$ monotonically increases with increasing Se concentration, which is different from the $x = 0.2$ case. The distinct behavior may originate from the FS topology, since theoretical studies propose that the superconducting gap structure and the $T_c$ change with the transition from the electron-pocket FSs ($x < 0.45$) around the $X$ points to hole-pocket FSs ($x > 0.45$) around the $\Gamma$ and $M$ points [56]. The localization state described as insulating or weakly insulating and WAL-WL-crossover or *weak* WAL regions (blue regions and green regions in Fig. 4) may relate to the CDW as discussed in the section above on the Hall coefficient, although we cannot completely determine the transition temperature of the CDW. However, we believe that the $T_{min}$ and $T^*$ are clues for revealing the physics where the localization states and superconductivity coexist in the BiCh$_2$-based compounds.

In conclusion, we have investigated the transport properties for the BiCh$_2$-based system LaO$_{1-x}$F$_x$BiS$_{2-y}$Se$_y$ ($x$ = 0.2 and 0.5, $y$ = 0–1.05). The $\rho_{ab}$ ($T$) behavior and $R_H$ ($T$) suggest the existence of unique localization sates. The localization behavior is systematically depressed with increasing Se concentration. MR measurements indicate that WAL is realized in the heavily Se-substituted systems. The WAL behavior is weakened by the changes in F and Se concentrations. A crossover state of WAL and WL emerges around moderately F-doped and Se-free LaO$_{0.8}$F$_{0.2}$BiS$_2$. The change of the MC behavior by the F and Se substitution clearly correlates with the results of $\rho_{ab}$ ($T$) and $R_H$ ($T$). Moreover, the localization regions of the WAL-WL crossover and *weak* WAL states are possibly associated with the CDW. On the basis of the results shown here, we propose that the BiCh$_2$-based system is a good platform to study unique localization states, including WAL, and its relation to SOC, electronic ordering, and superconductivity by using the elemental substitution with bulk single crystals.

**Materials and Methods**

The single crystals of LaO$_{1-x}$F$_x$BiS$_{2-y}$Se$_y$ were prepared by a high-temperature flux method by using a mixture of CsCl and KCl (CsCl:KCl = 5:3) as flux [59, 60]. We first synthesized polycrystalline samples by a conventional solid-state-reaction method [34, 35]. The obtained polycrystalline samples (0.4 g) and the flux (5.0 g) were mixed by a mortar and sealed into evacuated quartz tubes. The tubes were kept at 950 C° for 10 hours and slowly cooled to 600 C° at a rate of -2 C°/h. The heated tubes were opened in air and the obtained products were washed and filtered by pure water to remove the flux. The actual compositions of the obtained single crystals were estimated by energy-dispersive X-ray spectroscopy (EDX) with a scanning electron microscope (SEM) TM-3030 (Hitachi High-Tech). The Se concentration $y$ was estimated by the EDX. We use the nominal $x$-value in this study since the O and F concentration cannot be analyzed by the EDX due to their light mass. X-ray diffraction (XRD) was performed by a Miniflex600 (RIGAKU). The powder XRD patterns for polycrystalline samples were refined by the Rietveld analysis by using RIETAN-FP software [61] and the profiles for single crystalline samples were analyzed by PDIndexer software [62]. The lattice constants $a$ and $c$ for the polycrystalline and single crystals are displayed in *SI Appendix*, Fig. S1. The crystal structure image was prepared by VESTA [63].

The electrical resistivity, Hall resistivity, and magnetoresistance measurements on the single-crystalline samples were performed by a conventional four-terminal method in a physical properties measurement system (PPMS, Quantum Design) and a GM refrigerator (AXIS).

We performed first-principles band-structure calculation based on the density functional theory with the Perdew-Burke-Ernzerhof parametrization of the generalized gradient approximation [64] and the projector augmented wave (PAW) method [65] as implemented in the Vienna *ab initio* simulation package [66-69]. We used the experimental crystal structures determined by our experiment [70]. To represent the $O_{0.5}F_{0.5}$ occupation in $LaO_{0.5}F_{0.5}BiS_2$ and $LaO_{0.5}F_{0.5}BiSSe$, one oxygen and one fluorine atom are placed in a 10-atom unit cell. For simplicity, we assumed that selenium atoms occupy in-plane chalcogen sites in $LaO_{0.5}F_{0.5}BiSSe$. The core electrons in the PAW potential were $[Kr]4d^{10}$ for La, [He] for O and F, $[Xe]4f^{14}5d^{10}$ for Bi, [Ne] for S, and $[Ar]3d^{10}$ for Se, respectively. The spin-orbit coupling was included. We took 500 eV of the cutoff energy for the wave function, a 14×14×4 *k*-mesh, and 0.15 eV of the Gaussian smearing width.

ARPES measurements were performed at BL-9A in the Hiroshima Synchrotron Radiation Center (HiSOR) using a R4000 electron analyzer (Scienta Omicron) with *p*-polarized light. The total energy resolution was set to approximately 30 meV for $hv$ = 30 eV. All samples were cleaved *in-situ* on the (001) plane in an ultrahigh vacuum of less than $5 \times 10^{-9}$ Pa. All the measurements were performed at the same temperature of 13 K. The binding energies of the samples were determined by referencing the Fermi energy ($E_F$) of gold electronically contacted with samples.

**Acknowledgment**

K. Hoshi was supported by the Japanese Society for the Promotion of Science (JSPS) through a Research Fellowship for Young Scientists. This work was partly supported by JSPS KAKENHI (Grants Nos. JP18KK0076, JP20K20522, JP20J21627, JP21H00151, JP22K04908), and the Tokyo Government Advanced Research (Grant No. H31-1). The ARPES measurement was performed with the approval of the Proposal Assessing Committee of HSRC (Proposal No. 21BG033). We thank M. Arita and K. Shimada for support in the ARPES experiment.

Figure

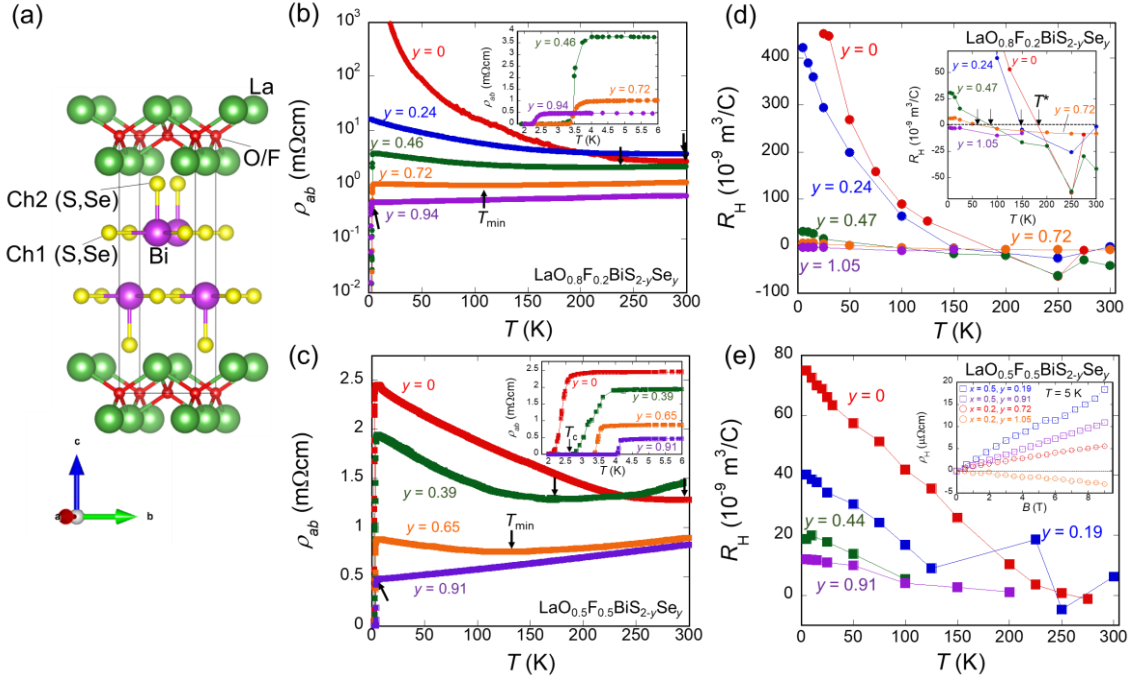

Fig. 1. (a) Schematic crystal structure of LaO$_{1-x}$F$_x$BiS$_{2-y}$Se$_y$. The solid line represents the unit cell. The tetragonal space group is *P*4/*nmm* (No. 129). (b,c) Temperature dependence of the electrical resistivity in the *ab*-plane, $\rho_{ab}$, for (b) $x = 0.2$ (circles) and (c) $x = 0.5$ (squares), respectively. Different colors refer to the actual Se concentration as indicated, estimated from EDX. The arrows in (b) and (c) show $T_{min}$ at which the normal-state resistivity above $T = 5$ K has a minimum. The insets in (b) and (c) represent the enlarged $\rho_{ab}$ (*T*) curves around the superconducting transition. The superconducting transition temperature $T_c$ is defined as the zero-resistivity temperature. (d,e) Temperature dependence of the Hall coefficient $R_H$ for (d) $x = 0.2$ (circles) and (e) $x = 0.5$ (squares), respectively. The inset of (d) expresses the enlarged $R_H$ (*T*) around the sign change. The arrows in the inset of (d) denote the sign-change temperature of the Hall coefficient. The inset of (e) shows the Hall resistivity $\rho_H$ as a function of the magnetic field at $T = 5$ K for LaO$_{0.5}$F$_{0.5}$BiS$_{1.81}$Se$_{0.19}$ (open blue squares), LaO$_{0.5}$F$_{0.5}$BiS$_{1.09}$Se$_{0.91}$ (open purple squares), LaO$_{0.8}$F$_{0.2}$BiS$_{1.28}$Se$_{0.72}$ (open red circles), and LaO$_{0.8}$F$_{0.2}$BiS$_{0.95}$Se$_{1.05}$ (open orange circles). Note that we plot the $x = 0.2$ data to compare with $x = 0.5$, although figure (e) represents the Hall coefficient for $x = 0.5$.

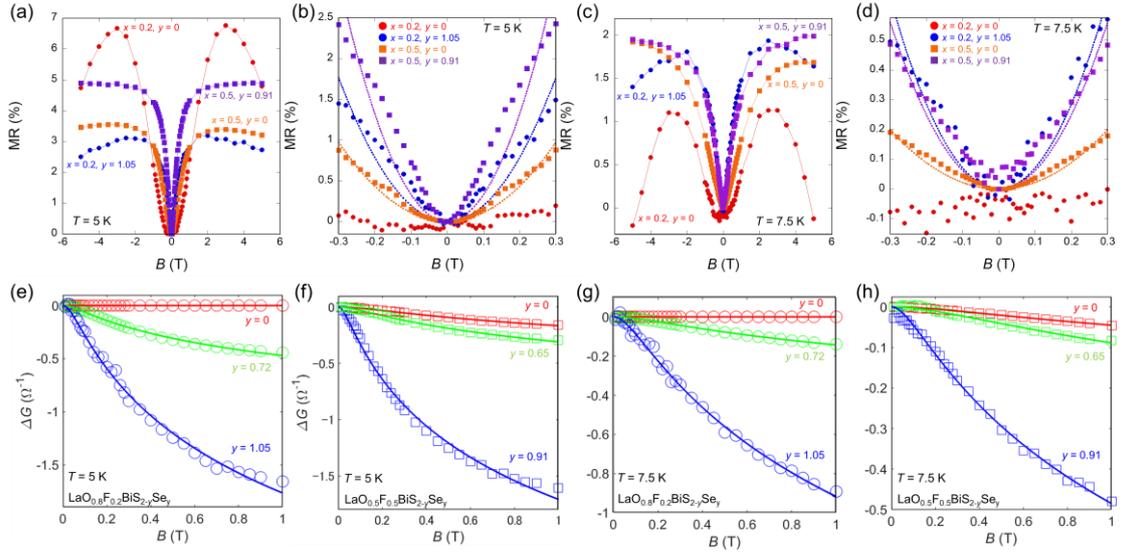

Fig. 2. (a) MR for $y = 0$ (red circles) and $y = 1.05$ (blue circles) with $x = 0.2$, and $y = 0$ (orange squares) and $y = 0.91$ (purple squares) with $x = 0.5$ at $T = 5$ K. (b) The enlarged MR curves in the low-field region at $T = 5.0$ K. (c) and (d) show the MR at $T = 7.5$ K. Dotted lines in (b) and (d) show the quadratic in $B$ fits. (e,f) MC at $T = 5$ K for (e) $x = 2$ (open circles) and (f) $x = 0.5$ (open squares), respectively. Solid lines represent the fits with the HLN model (Eq. 1). (g) and (h) express the MC at $T = 7.5$ K. The color variation expresses the actual Se concentration as listed, estimated from the EDX results. The MR curves for the LaO$_{0.8}$F$_{0.2}$BiS$_{1.28}$Se$_{0.72}$ ($x = 0.2$, $y = 0.72$) and LaO$_{0.5}$F$_{0.5}$BiS$_{1.35}$Se$_{0.65}$ ($x = 0.5$, $y = 0.65$) are displayed in *SI Appendix*, Fig. S2.

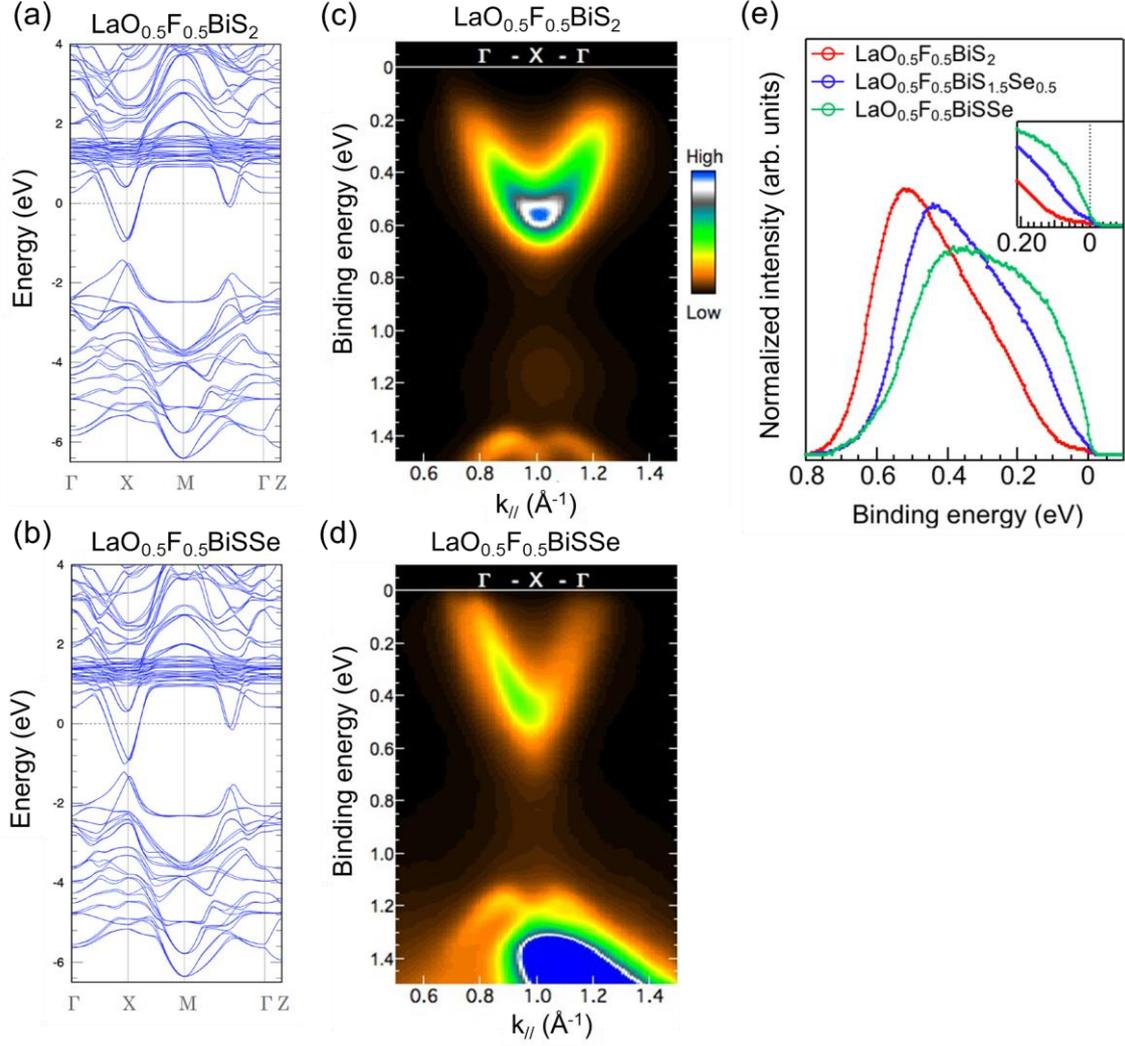

Fig. 3. Electronic band structures. (a,b) Calculated band structure of (a) LaO$_{0.5}$F$_{0.5}$BiS$_2$ ($x$ = 0.5, $y$ = 0) and (b) LaO$_{0.5}$F$_{0.5}$BiSSe ($x$ = 0.5, $y$ = 1). (c,d) ARPES spectra for the single crystals of (c) LaO$_{0.5}$F$_{0.5}$BiS$_2$ ($x$ = 0.5, $y$ = 0) and (d) LaO$_{0.5}$F$_{0.5}$BiSSe ($x$ = 0.5, $y$ = 1). (e) Comparison of the k-integrated spectra near $E_F$ for the electron pocket at $X$ between LaO$_{0.5}$F$_{0.5}$BiS$_2$, LaO$_{0.5}$F$_{0.5}$BiS$_{1.5}$Se$_{0.5}$, and LaO$_{0.5}$F$_{0.5}$BiSSe. The Se concentration for the ARPES results is given by its nominal value. The inset of (e) shows an expanded view of the $E_F$ region. The spectra were normalized by area after subtracting the Shirley background [57, 58].

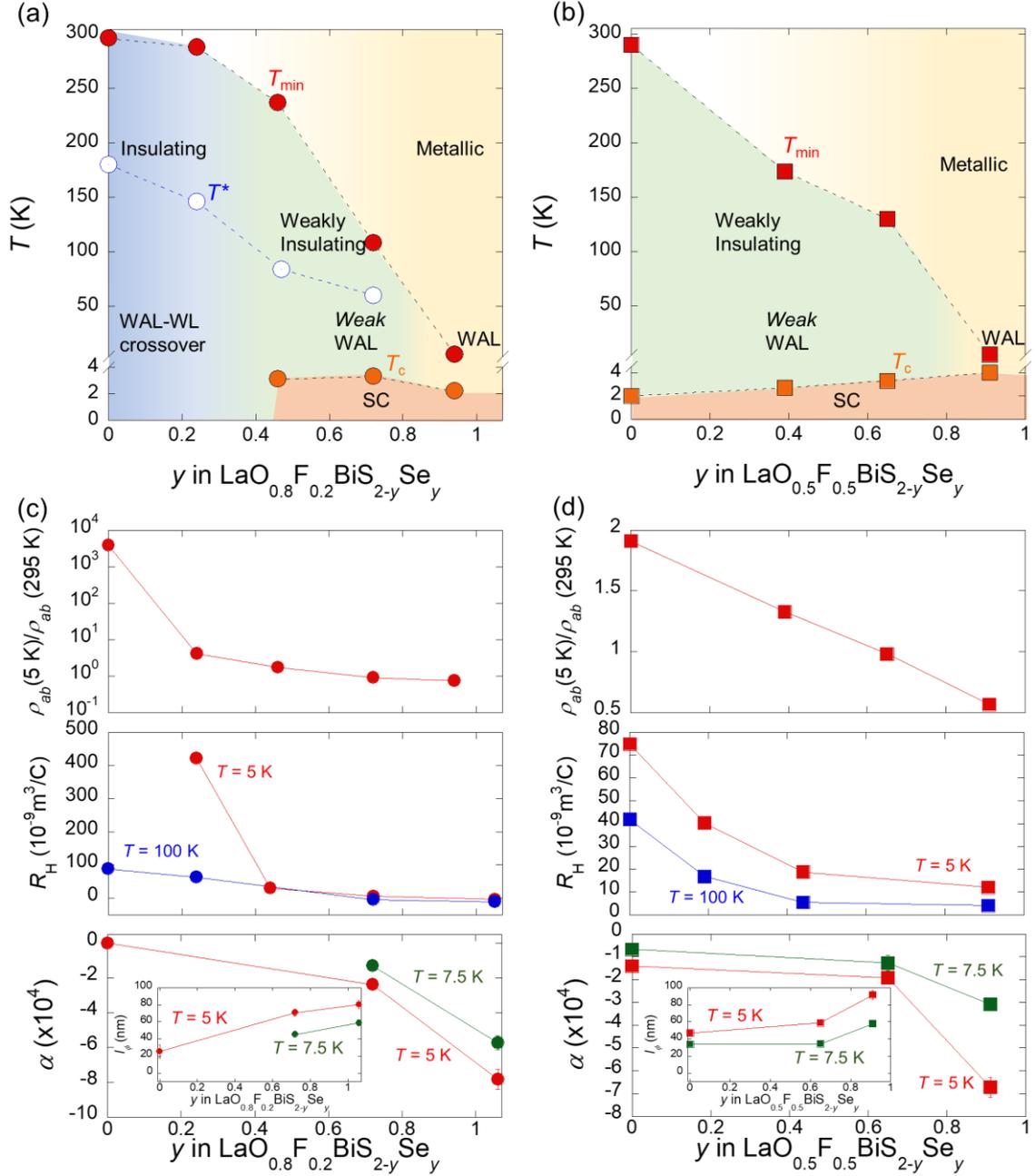

Fig. 4. Electronic phase diagrams for (a) $x = 0.2$ and (b) $x = 0.5$. $T_{min}$ (filled red circles) represents the temperature at which the $\rho_{ab}(T)$ above $T = 5$ K (in the normal states) has a minimum. $T^*$ (open blue circles) and $T_c$ (filled orange circles) exhibit the sign-change temperature in the Hall coefficient and superconducting transition temperature (zero resistivity). $T^*$ for $x = 0.5$ is not displayed in Fig. 4(b) since the sign change is only observed for $y = 0.19$. Blue, green, yellow, and orange regions express the insulating, weakly insulating, metallic, and superconducting (denoted as SC) states, respectively. We depict the WAL-WL crossover, *Weak* WAL, and WAL in (a) and (b) by referring to the results of the MC. (c,d) Se concentration $y$ dependence of the $\rho_{ab}$ (5 K)/$\rho_{ab}$ (295 K), Hall coefficients at $T = 5$ K

(red) and 100 K (blue), and $\alpha$ and $l_\phi$ [the inset in (c) and (d)] in the HLN model (Eq. 1) at $T = 5$ K (red) and 7.5 K (green) for (c) $x = 0.2$ (circles) and (d) $x = 0.5$ (squares), respectively.

Reference

[1] G. Bergmann, Influence of spin-orbit coupling on weak localization. *Phys. Rev. Lett.* **48**, 1046–1049 (1982).

[2] M. Aitani, T. Hirahara, S. Ichinokura, M. Hanaduka, D. Shin, and S. Hasegawa, *In situ* Magnetotransport Measurements in Ultrathin Bi Films: Evidence for Surface-Bulk Coherent Transport. *Phys. Rev. Lett.* **113**, 206802 (2014).

[3] M. Liu *et al.*, Crossover between Weak Antilocalization and Weak Localization in a Magnetically Doped Topological Insulator. *Phys. Rev. Lett.* **108**, 036805 (2012).

[4] Y. Pan *et al.*, Low carrier concentration crystals of the topological insulator $Bi_{2-x}Sb_xTe_{3-y}Se_y$: a magnetotransport study, *New J. Phys.* **16**, 123035 (2014).

[5] F. V. Tikhonenko, A. A. Kozikov, A. K. Savchenko, and R. V. Gorbachev, Transition between Electron Localization and Antilocalization in Graphene. *Phys. Rev. Lett.* **103**, 226801 (2009).

[6] H. Liu *et al.*, Quasi-2D Transport and Weak Antilocalization Effect in Few-layered $VSe_2$. *Nano Lett.* **19**, 4551–4559 (2019).

[7] S. Hikami, A. I. Larkin, and Y. Nagaoka, Spin-Orbit Interaction and Magnetoresistance in the Two Dimensional Random System. *Prog. Theor. Phys.* **63**, 707–710 (1980).

[8] H. Hayasaka and Y. Fuseya, Weak anti-localization in spin–orbit coupled lattice systems. *J. Phys.: Condens. Matter* **32**, 16LT01 (2020).

[9] Y. Mizuguchi *et al.*, $BiS_2$-based layered superconductor $Bi_4O_4S_3$. *Phys. Rev. B* **86**, 220510 (2012).

[10] Y. Mizuguchi, Material Development and Physical Properties of $BiS_2$-Based Layered Compounds. *J. Phys. Soc. Jpn.* **88**, 041001 (2019).

[11] Y. Mizuguchi *et al.*, Superconductivity in Novel $BiS_2$-Based Layered Superconductor $LaO_{1-x}F_xBiS_2$. *J. Phys. Soc. Jpn.* **81**, 114725 (2012).

[12] R. Higashinaka *et al.*, Pronounced −Log T Divergence in Specific Heat of Nonmetallic $CeOBiS_2$: A Mother Phase of $BiS_2$-Based Superconductor. *J. Phys. Soc. Jpn.* **84**, 023702 (2015).

[13] R. Jha, A. Kumar, S. K. Singh, and V. P. S. Awana, Synthesis and Superconductivity of New $BiS_2$ Based Superconductor $PrO_{0.5}F_{0.5}BiS_2$. *J. Supercond. Nov Magn.* **26**, 499–502 (2013).

[14] S. Demura *et al.*, New Member of $BiS_2$-Based Superconductor $NdO_{1-x}F_xBiS_2$. *J. Phys. Soc. Jpn.* **82**, 033708 (2013).

[15] Y. Kamihara, T. Watanabe, M. Hirano, and H. Hosono, Iron-Based Layered Superconductor

La[$O_{1-x}F_x$]FeAs ($x = 0.05-0.12$) with $T_c$ = 26 K. *J. Am. Chem. Soc.* **130**, 3296–3297 (2008).

[16] H. Usui, K. Suzuki, and K. Kuroki, Minimal electronic models for superconducting $BiS_2$ layers. *Phys. Rev. B* **86**, 220501 (2012).

[17] B. Li, Z. W. Xing and G. Q. Huang, Phonon spectra and superconductivity of the $BiS_2$-based compounds $LaO_{1-x}F_xBiS_2$. *Europhys. Lett.* **101**, 47002 (2013).

[18] K. Suzuki *et al.*, Electronic Structure and Superconducting Gap Structure in $BiS_2$-based Layered Superconductors. *J. Phys. Soc. Jpn.* **88**, 041008 (2019).

[19] M. Ochi, R. Akashi, and K. Kuroki, Strong Bilayer Coupling Induced by the Symmetry Breaking in the Monoclinic Phase of $BiS_2$-Based Superconductors. *J. Phys. Soc. Jpn.* **85**, 094705 (2016).

[20] M. A. Griffith, T. O. Puel, M. A. Continentino, and G. B. Martins, Multiband superconductivity in $BiS_2$-based layered compounds. *J. Phys.: Condens. Matter* **29**, 305601 (2017).

[21] G. B. Martins, A. Moreo, and E. Dagotto, RPA analysis of a two-orbital model for the $BiS_2$-based superconductors. *Phys. Rev. B* **87**, 081102 (2013).

[22] L. Jiao *et al.*, Evidence for nodeless superconductivity in $NdO_{1-x}F_xBiS_2$ ($x$ = 0.3 and 0.5) single crystals. *J. Phys.: Condens. Matter* **27**, 225701 (2015).

[23] N. Kase, Y. Terui, T. Nakano, and N. Takeda, Superconducting gap symmetry of the $BiS_2$-based superconductor $LaO_{0.5}F_{0.5}BiSSe$ elucidated through specific heat measurements. *Phys. Rev. B* **96**, 214506 (2017).

[24] T. Yamashita, Y. Tokiwa, D. Terazawa, M. Nagao, S. Watauchi, I. Tanaka, T. Terashima, and Y. Matsuda, Conventional *s*-Wave Superconductivity in $BiS_2$-Based $NdO_{0.71}F_{0.29}BiS_2$ Revealed by Thermal Transport Measurements. *J. Phys. Soc. Jpn.* **85**, 073707 (2016).

[25] K. Hoshi and Y. Mizuguchi, Experimental overview on pairing mechanisms of $BiCh_2$-based (Ch: S, Se) layered superconductors. *J. Phys.: Condens. Matter* **33**, 473001 (2021).

[26] Y. Ota *et al.*, Unconventional Superconductivity in the $BiS_2$-Based Layered Superconductor $NdO_{0.71}F_{0.29}BiS_2$. *Phys. Rev. Lett.* **118**, 167002 (2017).

[27] K. Hoshi, Y. Goto, and Y. Mizuguchi, Selenium isotope effect in the layered bismuth chalcogenide superconductor $LaO_{0.6}F_{0.4}Bi(S,Se)_2$. *Phys. Rev. B* **97**, 094509 (2018).

[28] R. Jha and Y. Mizuguchi, Unconventional isotope effect on transition temperature in $BiS_2$-based superconductor $Bi_4O_4S_3$. *Appl. Phys. Express* **13**, 093001 (2020).

[29] T. Machida *et al.*, "Checkerboard Stripe" Electronic State on Cleaved Surface of $NdO_{0.7}F_{0.3}BiS_2$ Probed by Scanning Tunneling Microscopy. *J. Phys. Soc. Jpn.* **83**, 113701 (2014).

[30] S. Demura, N. Ishida, Y. Fujisawa, and H. Sakata, Observation of Supermodulation in $LaO_{0.5}F_{0.5}BiSe_2$ by Scanning Tunneling Microscopy and Spectroscopy. *J. Phys. Soc. Jpn.* **86**, 113701 (2017).

[31] X. Zhang, Q. Liu, J.-W. Luo, A. J. Freeman, and A. Zunger, Hidden spin polarization in inversion-symmetric bulk crystals. *Nat. Phys.* **10**, 387–393 (2014).

[32] S.-L. Wu *et al*., Direct evidence of hidden local spin polarization in a centrosymmetric superconductor LaO$_{0.55}$F$_{0.45}$BiS$_2$. *Nat. Commun.* **8**, 1919 (2017).

[33] K. Hoshi, R. Kurihara, Y. Goto, M. Tokunaga, and Y. Mizuguchi, Extremely high upper critical field in BiCh$_2$-based (Ch: S and Se) layered superconductor LaO$_{0.5}$F$_{0.5}$BiS$_{2-x}$Se$_x$ ($x$ = 0.22 and 0.69). *Sci. Rep.* **12**, 288 (2022).

[34] T. Hiroi, J. Kajitani, A. Omachi, O. Miura, and Y. Mizuguchi, Evolution of Superconductivity in BiS$_2$-Based Superconductor LaO$_{0.5}$F$_{0.5}$Bi(S$_{1-x}$Se$_x$)$_2$. *J. Phys. Soc. Jpn.* **84**, 024723 (2015).

[35] K. Nagasaka *et al*., Intrinsic Phase Diagram of Superconductivity in the BiCh$_2$-Based System Without In-Plane Disorder. *J. Phys. Soc. Jpn.* **86**, 074701 (2017).

[36] H. Sakai *et al*., Insulator-to-Superconductor Transition upon Electron Doping in a BiS$_2$-Based Superconductor Sr$_{1-x}$La$_x$FBiS$_2$. *J. Phys. Soc. Jpn.* **83**, 014709 (2014).

[37] M. Tanaka *et al*., Site selectivity on chalcogen atoms in superconducting La(O,F)BiSSe. *Appl. Phys. Lett.* **106**, 112601 (2015).

[38] Y. Li *et al*., Electronic phase diagram in a new BiS$_2$-based Sr$_{1-x}$La$_x$FBiS$_2$ system. *Supercond. Sci. Technol.* **27**, 035009 (2014).

[39] J. Xing, S. Li, X. Ding, H. Yang, and H. H. Wen, Superconductivity appears in the vicinity of semiconducting-like behavior in CeO$_{1-x}$F$_x$BiS$_2$. *Phys. Rev. B* **86**, 214518 (2012).

[40] H. F. Zhai *et al*., Possible charge-density wave, superconductivity, and f-electron valence instability in EuBiS$_2$F. *Phys. Rev. B* **90**, 064518 (2014).

[41] J. Kajitani *et al*., Transverse-type Lattice Modulation in LaO$_{0.5}$F$_{0.5}$BiS$_2$: Possible Charge Density Wave Formation. *J. Phys. Soc. Jpn.* **90**, 103601 (2021).

[42] T. Yildirim, Ferroelectric soft phonons, charge density wave instability, and strong electron-phonon coupling in BiS$_2$ layered superconductors: A first-principles study. *Phys. Rev. B* **87**, 020506 (2013).

[43] X. Wan, H.-C. Ding, S. Y. Savrasov, and C.-G. Duan, Electron-phonon superconductivity near charge-density-wave instability in LaO$_{0.5}$F$_{0.5}$BiS$_2$: Density-functional calculations. *Phys. Rev. B* **87**, 115124 (2013).

[44] H. Fukuyama and K. Yoshida, Negative Magnetoresistance in the Anderson Localized States. *J. Phys. Soc. Jpn.* **46**, 102–105 (1979).

[45] W. Bai *et al*., Intrinsic Negative Magnetoresistance in Van Der Waals FeNbTe$_2$ Single Crystals. *Adv Mater.* **31**, 1900246 (2019).

[46] K. Terashima *et al*., Proximity to Fermi-surface topological change in superconducting LaO$_{0.54}$F$_{0.46}$BiS$_2$. *Phys. Rev. B* **90**, 220521 (2014).

[47] O. Pavlosiuk *et al*., Antiferromagnetism and superconductivity in the half-Heusler semimetal HoPdBi. *Sci. Rep.* **6**, 18797 (2016).

[48] S. Hao *et al*., Low-Temperature Eutectic Synthesis of PtTe$_2$ with Weak Antilocalization and

Controlled Layer Thinning. *Adv. Funct. Mater.* **28**, 1803746 (2018).

[49] Z. Hou, *et al.*, Large low-field positive magnetoresistance in nonmagnetic half-Heusler ScPtBi single crystal. *Appl. Phys. Lett.* **107**, 202103 (2015).

[50] Y. Ma, Y. Wang, and G. Wang, Anisotropic magnetoresistance and possible weak antilocalization in $Mg_3Bi_2$ single crystal. *Europhys. Lett.* **138**, 36003 (2022).

[51] Q. Liu *et al.*, Search and design of nonmagnetic centrosymmetric layered crystals with large local spin polarization. *Phys. Rev. B* **91**, 235204 (2015).

[52] H. Huang *et al.*, Enhancing thermoelectric performance of $SrFBiS_{2-x}Se_x$ via band engineering and structural texturing. *J. Materiomics.* **8**, 302–310 (2022).

[53] R. Sagayama *et al.*, Symmetry Lowering in $LaOBiS_2$: A Mother Material for $BiS_2$-Based Layered Superconductors. *J. Phys. Soc. Jpn.* **84**, 123703 (2015).

[54] N. Hirayama, M. Ochi, and K. Kuroki, Theoretical study of fluorine doping in layered $LaOBiS_2$-type compounds. *Phys. Rev. B* **100**, 125201 (2019).

[55] K. Hoshi *et al.*, Structural Phase Diagram of $LaO_{1-x}F_x$BiSSe: Suppression of the Structural Phase Transition by Partial F Substitutions. *Condensed Matter*, **5**, 81 (2020).

[56] T. Agatsuma and T. Hotta, Fermi-surface topology and pairing symmetry in $BiS_2$-based layered superconductors. *J. Magn. Magn. Mater.* **400**, 73–80 (2016).

[57] D. A. Shirley, High-Resolution X-Ray Photoemission Spectrum of the Valence Bands of Gold. *Phys. Rev. B* **5**, 4709–4714 (1972).

[58] A. Proctor and P. M. A. Sherwood, Data Analysis Techniques in X-ray Photoelectron Spectroscopy. *Anal. Chem.* **54**, 13–19 (1982).

[59] M. Nagao, A. Miura, S. Demura, K. Deguchi, S. Watauchi, T. Takei, Y. Takano, N. Kumada, and I. Tanaka, Growth and superconducting properties of F-substituted $ROBiS_2$ (R=La, Ce, Nd) single crystals. *Solid State Commun.* **178**, 33–36 (2014).

[60] M. Nagao, Growth and characterization of $R(O,F)BiS_2$ (R = La, Ce, Pr, Nd) superconducting single crystals. *Novel Supercond. Mater.* **1**, 64–74 (2015).

[61] F. Izumi and K. Momma, Three-Dimensional Visualization in Powder Diffraction, *Solid State Phenom.* **130**, 15–20 (2007).

[62] Y. Seto, D. Nishio-Hamane, T. Nagai, and N. Sata, Development of a Software Suite on X-ray Diffraction Experiments. *Review of High Pressure Science and Technology* **20**, 269–276 (2010), (in Japanese).

[63] K. Momma and F. Izumi, VESTA 3 for three-dimensional visualization of crystal, volumetric and morphology data. *J. Appl. Crystallogr.* **44**, 1272–1276 (2011).

[64] J. P. Perdew, K. Burke, and M. Ernzerhof, Generalized Gradient Approximation Made Simple. *Phys. Rev. Lett.* **77**, 3865–3868 (1996).

[65] G. Kresse and D. Joubert, From ultrasoft pseudopotentials to the projector augumented-wave

method. *Phys. Rev. B* **59**, 1758–1775 (1999).

[66] G. Kresse and J. Hafner, *Ab initio* molecular dynamics for liquid metals. *Phys. Rev. B* **47**, 558–561 (1993).

[67] G. Kresse and J. Hafner, *Ab initio* molecular-dynamics simulation of the liquid-metal-amorphous-semiconductor transition in germanium. *Phys. Rev. B* **49**, 14251–14269 (1994).

[68] G. Kresse and J. Furthmüller, Efficiency of ab-initio total energy calculations for metals and semiconductors using a plane-wave basis set. *Comput. Mater. Sci.* **6**, 15–50 (1996).

[69] G. Kresse and J. Furthmüller, Efficient iterative schemes for ab initio total-energy calcilations using a plane-wave basis set. *Phys. Rev. B* **54**, 11169–11186 (1996).

[70] Y. Mizuguchi *et al*., In-plane chemical pressure essential for superconductivity in BiCh$_2$-based (Ch: S, Se) layered structure. *Sci. Rep.* **5**, 14968 (2015).

**Supporting Information**

Weak antilocalization induced by Se substitution in layered BiCh$_2$-based (Ch = S, Se) superconductors LaO$_{1-x}$F$_x$BiS$_{2-y}$Se$_y$

K. Hoshi[1], H. Arima[1], N. Kataoka[2], M. Ochi[3,4], A. Yamashita[1], A. de Visser[5], T. Yokoya[2,6], K. Kuroki[4] and Y. Mizuguchi[1*]

[1]Department of Physics, Tokyo Metropolitan University, Hachioji 192-0397, Japan
[2]Graduate School of Natural Science and Technology, Okayama University, Okayama 700-8530, Japan
[3]Forefront Research Center, Osaka University, Toyonaka, Osaka 560-0043, Japan
[4]Department of Physics, Osaka University, Toyonaka, Osaka 560-0043, Japan
[5]Van der Waals–Zeeman Institute, University of Amsterdam, Science Park 904, 1098 XH Amsterdam, The Netherlands
[6]Research Institute for Interdisciplinary Science, Okayama University, Okayama 700-8530, Japan

S1. Lattice constants

We performed X-ray diffraction (XRD) measurements for polycrystalline samples and single crystals of BiCh$_2$-based LaO$_{1-x}$F$_x$BiS$_{2-y}$Se$_y$ and analyzed the XRD patterns to estimate the lattice constants. The refined lattice constants $a$ and $c$ for the polycrystalline samples and single crystals are plotted in Fig. S1 (a) and (b). Note that we present only the $c$-axis lattice constant for the single crystals because only the 00$l$ peaks were used for the refinement. The $a$-axis lattice parameter increases with increasing Se concentration since the ionic radius of Se is larger than that of S. The $c$-axis lattice constant of $x = 0.5$ is smaller than of $x = 0.2$ since the electron-carrier doping (F substitution) is greatly sensitive to changing the $c$-axis length in the BiCh$_2$-based system [1, 2]. Thus, there are slight variations of the $c$-axis lengths with the Se concentration because the lattice constant $c$ is dependent on both F and Se concentration. The trend of the $c$-axis parameters of the single crystals with the Se concentration is in good agreement with that of the polycrystalline samples. Moreover, the actual compositions from the EDX results are nearly consistent with the nominal compositions. The results indicate that the F and Se concentrations are successfully controlled in our single crystals.

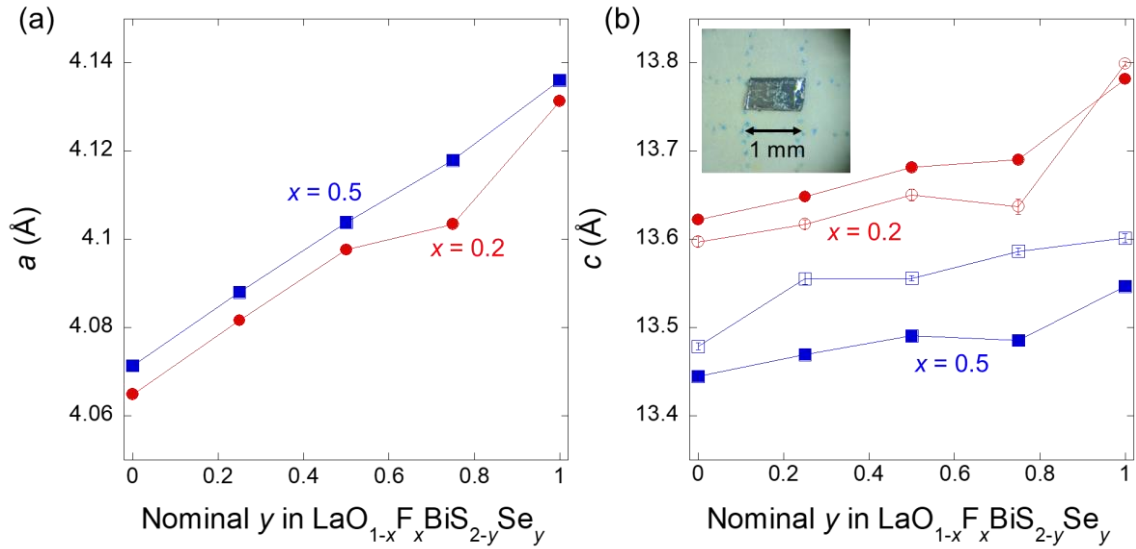

Fig. S1. Lattice constants (a) $a$ and (b) $c$ as a function of the nominal Se concentration $y$ for the polycrystalline samples (closed red circles for $x = 0.2$ and blue squares for $x = 0.5$) and single crystals (open red circles for $x = 0.2$ and blue squares for $x = 0.5$) of the LaO$_{1-x}$F$_x$BiS$_{2-y}$Se$_y$. The inset of (b) shows a photo of a single crystal of LaO$_{0.5}$F$_{0.5}$BiS$_{1.61}$Se$_{0.39}$ as a representative of the LaO$_{1-x}$F$_x$BiS$_{2-y}$Se$_y$ series.

S2. Magnetoresistance measurements for moderately Se-substituted $LaO_{0.5}F_{0.5}BiS_{1.35}Se_{0.65}$ and $LaO_{0.8}F_{0.2}BiS_{1.28}Se_{0.72}$

We display magnetoresistance (MR) data for moderately Se-substituted $LaO_{0.8}F_{0.2}BiS_{1.28}Se_{0.72}$ ($x = 0.2$, $y = 0.72$) and $LaO_{0.5}F_{0.5}BiS_{1.35}Se_{0.65}$ ($x = 0.5$, $y = 0.72$) at $T = 5$ K [Figs. S2(a) and (b)] and $T = 7.5$ K [Figs. S2(c) and (d)]. The low-field MR curves at $T = 5$ K have a cusp-like behavior, which indicates weak antilocalization (WAL) states. However, the MR is broadened compared with the heavily Se-substituted compounds [Fig. 2(b)]. Thus, we deduce that the moderately Se-substituted systems are in *weak* WAL states.

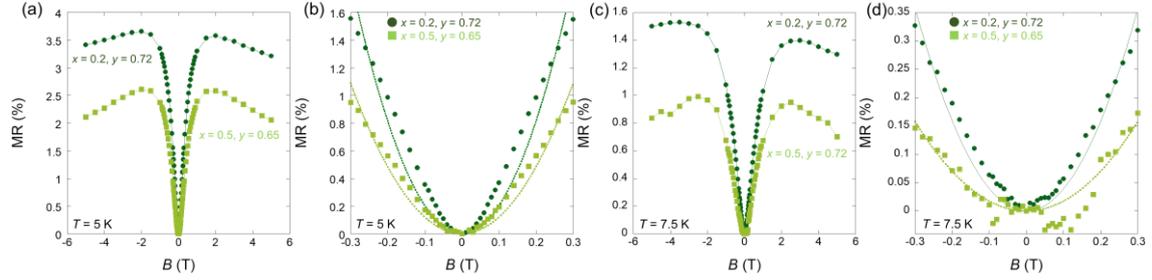

Fig. S2. MR for the $LaO_{0.8}F_{0.2}BiS_{1.28}Se_{0.72}$ (dark green circles) and $LaO_{0.5}F_{0.5}BiS_{1.35}Se_{0.65}$ (pale green squares) at (a) $T = 5$ K and (c) $T = 7.5$ K. The enlarged MR curves in the low-field region at (b) $T = 5.0$ K and (d) $T = 7.5$ K. Dotted lines in (b) and (d) show the quadratic fits. The actual Se concentration was estimated by EDX.

## S3. Magnetization measurements

We performed magnetization measurements to estimate the shielding volume fractions for the LaO$_{1-x}$F$_x$BiS$_{2-y}$Se$_y$ samples. Figures S3(a) and (b) show the temperature dependence of the magnetic susceptibility under $H$ = 1 Oe for $x$ = 0.2 and 0.5, respectively. The demagnetization factor $N$ is large in this study since our single crystals have a plate-like shape and the applied magnetic field is parallel to the $c$ axis (perpendicular to the plate surface) [3, 4]. Thus, we corrected the diamagnetic signal by considering the demagnetization effect and the estimated $N$ is displayed in Table S1. The observed $T_c$ is almost consistent with the resistivity measurements [see Fig. 1(b) and (c) in main text]. The shielding fraction after demagnetization correction increase with increasing Se concentration [see Fig. S3(c)], which is comparable to the results obtained on of polycrystals [5]. The magnetic susceptibility for LaO$_{0.8}$F$_{0.2}$BiS$_{1.28}$Se$_{0.72}$, LaO$_{0.5}$F$_{0.5}$BiS$_{1.35}$Se$_{0.65}$, and LaO$_{0.5}$F$_{0.5}$BiS$_{1.06}$Se$_{0.94}$ slightly exceeds the perfect diamagnetism ($4\pi\chi$ = -1) probably because of the existence of tiny single crystals on the main single-crystalline sample. These results demonstrate that the Se substitution is essential for bulk superconductivity to emerge, which supports the scenario that the metallic behavior in the entire temperature regions and WAL states emerge in the heavily Se-substituted compounds, and it is suppressed with decreasing Se concentration. (see main text).

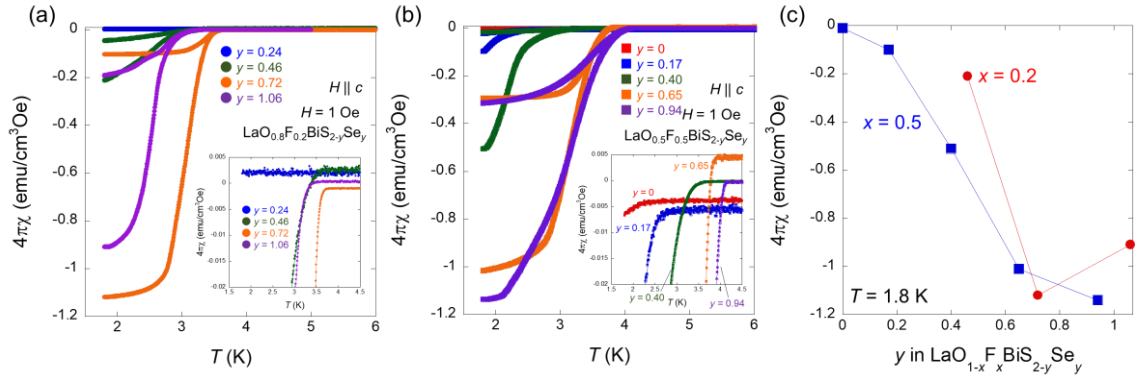

Fig. S3. Temperature dependence of the magnetic susceptibility after demagnetization correction for (a) $x$ = 0.2 and (b) $x$ = 0.5, respectively. The applied magnetic field is parallel to the $c$ axis. Color difference denote the actual Se concentration estimated from EDX. The inset of (a) and (b) show enlarged magnetic-susceptibility curves around the superconducting transition. (c) Magnetic susceptibility at $T$ = 1.8 K as a function of the Se concentration $y$ for $x$ = 0.2 (red circles) and $x$ = 0.5 (blue squares).

Table S1. $T_c$, samples size, demagnetization factor $N$, and magnetic susceptibility of $LaO_{1-x}F_xBiS_{2-y}Se_y$. The Se concentration is estimated from the EDX.

| Compounds | $T_c$ from magnetization curve (K) | Sample size ($a \times b \times c$) (mm$^3$) | $N$ | $4\pi\chi$ at $T = 1.8$ K |
|---|---|---|---|---|
| $LaO_{0.8}F_{0.2}BiS_{1.76}Se_{0.24}$ | - | $1.025 \times 0.553 \times 0.169$ | 0.706 | - |
| $LaO_{0.8}F_{0.2}BiS_{1.54}Se_{0.46}$ | 3.5 | $1.146 \times 0.448 \times 0.063$ | 0.858 | -0.21 |
| $LaO_{0.8}F_{0.2}BiS_{1.28}Se_{0.72}$ | 3.7 | $1.132 \times 1.812 \times 0.126$ | 0.849 | -1.12 |
| $LaO_{0.8}F_{0.2}BiS_{0.94}Se_{1.06}$ | 3.4 | $0.984 \times 0.950 \times 0.154$ | 0.760 | -0.91 |
| $LaO_{0.5}F_{0.5}BiS_2$ | 2.2 | $0.919 \times 0.867 \times 0.149$ | 0.752 | -0.01 |
| $LaO_{0.5}F_{0.5}BiS_{1.83}Se_{0.17}$ | 2.6 | $0.628 \times 0.739 \times 0.053$ | 0.861 | -0.1 |
| $LaO_{0.5}F_{0.5}BiS_{1.61}Se_{0.39}$ | 3.4 | $1.292 \times 1.472 \times 0.109$ | 0.858 | -0.51 |
| $LaO_{0.5}F_{0.5}BiS_{1.35}Se_{0.65}$ | 3.9 | $1.087 \times 0.582 \times 0.085$ | 0.827 | -1.01 |
| $LaO_{0.5}F_{0.5}BiS_{1.06}Se_{0.94}$ | 4.1 | $1.038 \times 1.158 \times 0.062$ | 0.892 | -1.14 |

S4. Fitting parameters in the HLN formula

We use the 2D Hikami-Larkin-Nagaoka (HLN) formula (Eq. 1 in the main text) [6] to fit the MC data for $LaO_{1-x}F_xBiS_{2-y}Se_y$. There are two fitting parameters in the HLN model: the prefactor $\alpha$ and phase coherence length $l_\phi$. These parameters and the coefficient of determination $R^2$ in the HLN fitting are summarized in Table S2. Over 95% of $R^2$ is obtained except for $LaO_{0.8}F_{0.2}BiS_2$ at $T = 7.5$ K, which exhibits that the MC curves are well-fitted with the HLN formula. The $R^2$ value for $LaO_{0.8}F_{0.2}BiS_2$ at $T = 7.5$ K is roughly 75% probably due to the WAL-WL crossover state for this compound. The fact that the obtained value of $\alpha$ for $LaO_{0.8}F_{0.2}BiS_2$ is much smaller than the others can provide the evidence of WAL-WL crossover state for this compound. Se substitution leads to an increase in the absolute value of $\alpha$ and $l_\phi$, which indicates that the WAL state emerges by the Se substitution. We plot the Se-concentration dependence of these parameters in Figs. 4(c) and (d) in the main text.

Table S2. HLN parameters $\alpha$ and phase coherence length $l_\phi$ and coefficient of determination $R^2$ in the HLN fitting for $LaO_{1-x}F_xBiS_{2-y}Se_y$ at $T = 5$ K and 7.5 K. The Se concentration is estimated from the EDX.

| Compounds | Temperature (K) | $\alpha$ (x$10^4$) | $l_\phi$ (nm) | $R^2$ |
|---|---|---|---|---|
| $LaO_{0.8}F_{0.2}BiS_2$ | 5 | $-0.000968\pm0.000791$ | $25.3\pm7.6$ | 0.953 |
| $LaO_{0.8}F_{0.2}BiS_{1.28}Se_{0.72}$ | 5 | $-2.38\pm0.16$ | $70.7\pm3.6$ | 0.996 |
| $LaO_{0.8}F_{0.2}BiS_{0.95}Se_{1.05}$ | 5 | $-7.85\pm0.57$ | $80.2\pm4.6$ | 0.994 |
| $LaO_{0.5}F_{0.5}BiS_2$ | 5 | $-1.40\pm0.08$ | $46.9\pm1.6$ | 0.998 |
| $LaO_{0.5}F_{0.5}BiS_{1.35}Se_{0.65}$ | 5 | $-1.92\pm0.13$ | $58.8\pm2.6$ | 0.997 |
| $LaO_{0.5}F_{0.5}BiS_{1.09}Se_{0.91}$ | 5 | $-6.72\pm0.44$ | $91.5\pm5.1$ | 0.995 |
| $LaO_{0.8}F_{0.2}BiS_2$ | 7.5 | $-0.00020\pm0.015$ | $14.3\pm30.3$ | 0.752 |
| $LaO_{0.8}F_{0.2}BiS_{1.28}Se_{0.72}$ | 7.5 | $-1.29\pm0.05$ | $45.0\pm1.0$ | 0.999 |
| $LaO_{0.8}F_{0.2}BiS_{0.95}Se_{1.05}$ | 7.5 | $-5.73\pm0.40$ | $58.7\pm2.7$ | 0.997 |
| $LaO_{0.5}F_{0.5}BiS_2$ | 7.5 | $-0.67\pm0.06$ | $33.7\pm0.1$ | 0.995 |
| $LaO_{0.5}F_{0.5}BiS_{1.35}Se_{0.65}$ | 7.5 | $-1.28\pm0.33$ | $34.3\pm4.1$ | 0.985 |
| $LaO_{0.5}F_{0.5}BiS_{1.09}Se_{0.9}$ | 7.5 | $-3.09\pm0.26$ | $57.9\pm3.1$ | 0.998 |

Reference

[1] K. Nagasaka *et al*., Intrinsic Phase Diagram of Superconductivity in the BiCh$_2$-Based System Without In-Plane Disorder. *J. Phys. Soc. Jpn.* **86**, 074701 (2017).

[2] Y. Mizuguchi, Material Development and Physical Properties of BiS$_2$-Based Layered Compounds. *J. Phys. Soc. Jpn.* **88**, 041001 (2019).

[3] E. H. Brandt, Irreversible magnetization of pin-free type-II superconductors. *Phys. Rev. B* **60**, 11939–11942 (1999).

[4] M. Abdel-Hafiez *et al*., Temperature dependence of lower critical field H$_{c1}$(*T*) shows nodeless superconductivity in FeSe. *Phys. Rev. B* **88**, 174512 (2013).

[5] T. Hiroi, J. Kajitani, A. Omachi, O. Miura, and Y. Mizuguchi, Evolution of Superconductivity in BiS$_2$-Based Superconductor LaO$_{0.5}$F$_{0.5}$Bi(S$_{1-x}$Se$_x$)$_2$. *J. Phys. Soc. Jpn.* **84**, 024723 (2015).

[6] S. Hikami, A. I. Larkin, and Y. Nagaoka, Spin-Orbit Interaction and Magnetoresistance in the Two Dimensional Random System. *Prog. Theor. Phys.* **63**, 707–710 (1980).